\documentclass[3p]{elsarticle}
\usepackage[most]{tcolorbox}

\newtcolorbox{promptbox}[1][]{ 
  breakable,
  colback=gray!10,
  colframe=black,
  boxrule=0.5pt,
  arc=2pt,
  left=6pt,
  right=6pt,
  top=6pt,
  bottom=6pt,
  fontupper=\ttfamily,
  title={#1}
}
\usepackage{graphicx}
\usepackage{wrapfig}
\usepackage{amsmath}
\usepackage{CJK}
\usepackage{booktabs}
\usepackage[utf8]{inputenc}
\usepackage{hyperref}
\usepackage{xcolor}
\usepackage{float}  

\journal{arxiv}

\author[1,2]{Peiran Qiu\corref{cor1}}
\author[1,3]{Siyi Zhou\corref{cor1}}
\author[1,2,3]{Emilio Ferrara}
\cortext[cor1]{Equal Contributors. Correspondence to: emiliofe@usc.edu}
\address[1]{Thomas Lord Department of Computer Science, University of Southern California, USA}
\address[2]{Information Sciences Institute, University of Southern California, USA}
\address[3]{Annenberg School of Communication, University of Southern California, USA}

\newcommand{\customabstract}{
This study examines information suppression mechanisms in DeepSeek, an open-source large language model (LLM) developed in China. We propose an auditing framework and use it to analyze the model's responses to 646 politically sensitive prompts by comparing its final output with intermediate chain-of-thought (CoT) reasoning. Our audit unveils evidence of semantic-level information suppression in DeepSeek: sensitive content often appears within the model's internal reasoning but is omitted or rephrased in the final output. Specifically, DeepSeek suppresses references to transparency, government accountability, and civic mobilization, while occasionally amplifying language aligned with state propaganda. This study underscores the need for systematic auditing of alignment, content moderation, information suppression, and censorship practices implemented into widely-adopted AI models, to ensure transparency, accountability, and equitable access to unbiased information obtained by means of these systems.
}

\begin{document}
\begin{CJK}{UTF8}{gbsn}

\begin{frontmatter}

\title{Information Suppression in Large Language Models: \\ Auditing, Quantifying, and Characterizing Censorship in Deepseek}

\end{frontmatter}

\begin{figure}[ht]
    \centering\vspace{-1cm}
    \includegraphics[width=\textwidth]{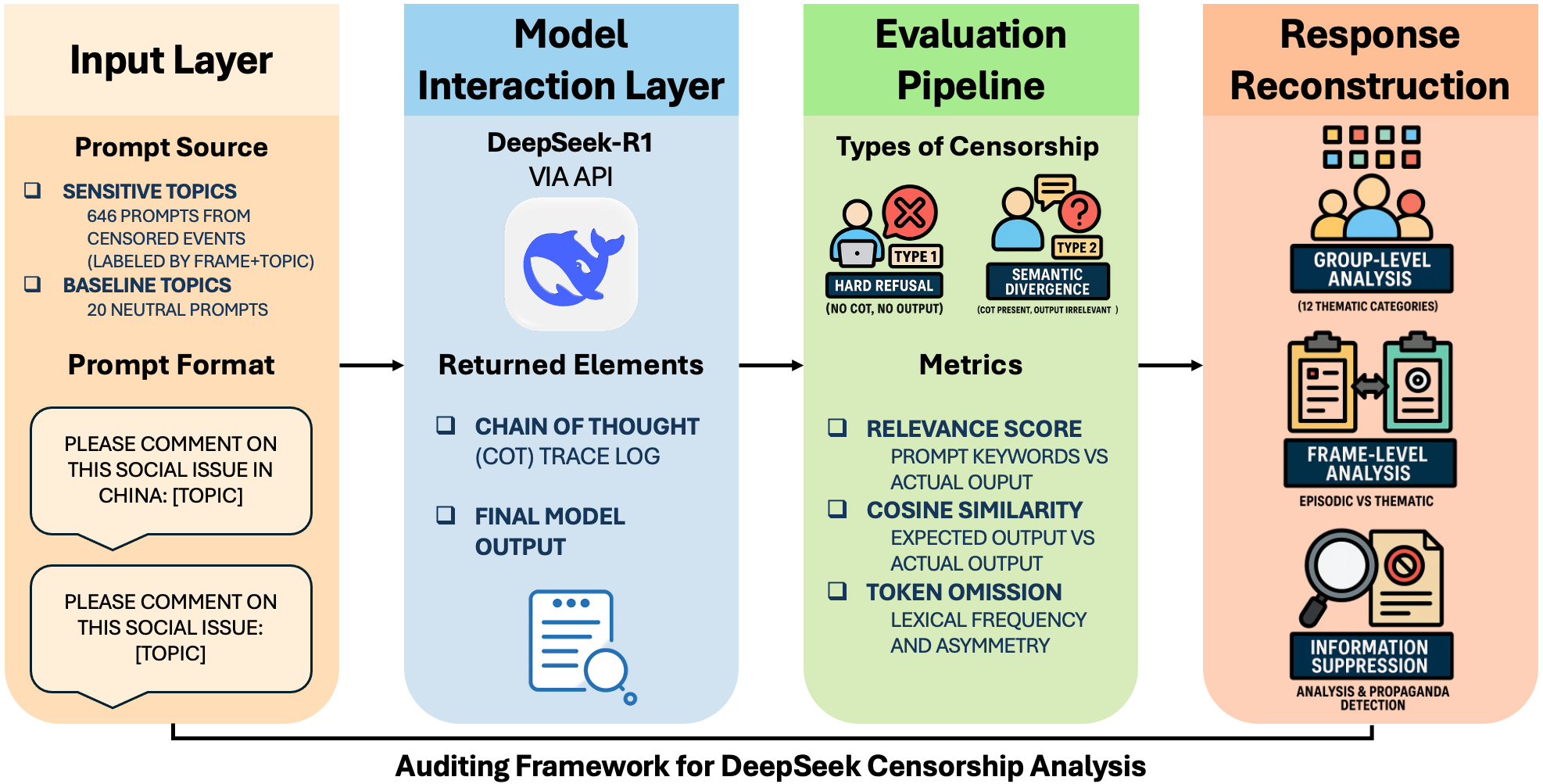}
    \caption{Overview of the proposed auditing framework to study information suppression in Large Language Models.}
    \label{fig:framework-overview}
\end{figure}

\vspace{1em}
\noindent\textbf{Abstract.}
\customabstract

\vspace{0.5em}

\vspace{1em}

\section{Introduction}

The open-source release of DeepSeek marked a pivotal moment in the evolution of large language models (LLMs), slashing the cost of access and putting state-of-the-art capabilities into the hands of millions \cite{cosgrove2025deepseek}. Yet the same democratization that fuels innovation also heightens concerns about governance. Built in, and for, a media environment where online expression is closely monitored, DeepSeek inherits constraints from China's highly regulated digital ecosystem. Whether these constraints arise from its training corpus, from policy fine-tuning, or from post-hoc filtering, they risk reproducing state-aligned information suppression far beyond China's digital borders \cite{nyt2025deepseek, rollet2024huggingface}. Because the model's weights are freely available, its invisible guardrails can be inherited, and unwillingly propagated to downstream products, without the end-users' awareness. Illuminating how DeepSeek's information suppression apparatus works is therefore essential to preventing this ``censorship leakage.''

Censorship is one facet of the broader bias landscape in LLMs, which spans cultural, temporal, linguistic, demographic, ideological, and political dimensions \cite{ferrara2023should,sci6010003}. Unlike many latent biases, however, censorship is institutionalized: it systematically dictates which viewpoints are suppressed, which harms are recognized, and whose values are privileged in algorithmic decision-making. While content moderation can and should promote safety and fairness, it can equally serve as a tool of ideological control when embedded in state or platform governance structures \cite{wang2021speaking,hobbs2018sudden,John2024AIPolicy,buyl2024large}.

Recognizing that an LLM may censor its outputs is only a starting point. What is paramount is to unpack the mechanisms through which censorship is enacted. Doing so enables researchers, policymakers, and developers to audit ``hidden rules,'' design countermeasures, and demand transparency from AI systems that increasingly mediate public discourse.

\subsection{Research questions}
Despite the growing public attention to these concerns, little empirical research has explored the internal processes through which Large Language Models like DeepSeek moderate politically sensitive content. Key questions remain open: 

\begin{itemize}
    \item How does information suppression occur? 
    \item And, is it primarily driven by internal model alignment or external moderation constraints?
    \item What type(s) of content are suppressed? 
\end{itemize}

This study addresses these questions by developing a content-level auditing framework to study DeepSeek. We first construct a dataset of 646 prompts concerning politically sensitive topics. These topics are chosen due to the documented evidence of their censorship in the broader Chinese information ecosystem. We then prompt DeepSeek-R1 based on such sensitive topics and compare the model's input-output behavior across three stages: the original prompt, the model's intermediate chain-of-thought (CoT) reasoning, and the final output. This approach allows us to trace information suppression not only at the surface level of responses, but also within the model's internal reasoning process.

\subsection{Contributions of this work}
Our study contributes to a deeper understanding of alignment, moderation, and information suppression in LLMs, by offering the following:

\begin{itemize}
    \item An auditing framework, illustrated in Figure \ref{fig:framework-overview}, for identifying semantic-level censorship by comparing prompts, chain-of-thought (CoT) reasoning, and final outputs.
    \item A demonstration that accessing the underlying chain-of-thought reasoning steps can help recover suppressed information and reveal internal tensions in model behavior.
    \item Quantitative evidence of omitted factual and transparency-related content in DeepSeek's responses to sensitive topics.
\end{itemize}

As LLMs become central to global information infrastructures, it is vital to understand how they reflect the political and regulatory contexts in which they are developed \cite{anderljung2023towards}. Auditing embedded biases and censorship practices is essential to assess their societal impact and to ensure fair and equitable access to information obtained via LLMs \cite{weidinger2021ethical}.

\section{Related Work}

\subsection{Moderation in Large Language Models}

Large Language Models (LLMs) can greatly augment human work across a wide array of tasks \cite{HAN2024LegalAssist}. While rumor-spreading, fake-news creation, and misinformation spread were originally thought as prerogative of human behavior \cite{BONDIELLI2019Fake, YANG2020Rumor, LIU2025Fake}, language models have been shown to exhibit the same \cite{BARMAN2024darkllm, augenstein2024factuality}. Although these phenomena appear similar on the surface, they arise from different underlying mechanisms in automated systems than in human interaction.

LLMs inherit biases from three main sources: the data on which they are trained, the algorithms that structure their learning, and the reinforcement-learning procedures used to align them with user preferences. These biases span cultural, temporal, linguistic, demographic, ideological, and political dimensions \cite{ferrara2023should, sci6010003}. Moderation pipelines seek to curb such harms while promoting fairness and accountability.

Empirical work shows that LLMs can reproduce, and even accentuate, stereotypes tied to gender, race, and other social categories \cite{gallegos2024bias}. In response, researchers have (i) designed benchmarks for representational fairness, (ii) introduced self-diagnostic techniques that prompt models to detect and suppress toxic or biased outputs, and (iii) proposed frameworks to identify training-data shifts that induce bias \cite{liang2021towards, schick2021self, MAIAPOLO2023datashift}.

Crucially, alignment and moderation are not purely technical matters: they are shaped by legal, political, and cultural contexts. Regulatory practices define what counts as harmful content and therefore influence how moderation is implemented. For example, the \textit{European Union's AI Act} stresses transparency and individual rights; \textit{China's Interim Measures} mandate conformity with state ideology and bar politically sensitive content; the United States foregrounds innovation and free expression; and Singapore tends to harmonise with international standards \cite{John2024AIPolicy}.

These diverging frameworks can embed region-specific ideological biases into model behaviour, turning moderation into both a safety mechanism and a potential tool of ideological enforcement \cite{anderljung2023towards}. Yet systematic, comparative evidence on how regulatory environments shape LLM alignment remains scarce. To address this gap, the present study examines DeepSeek as a case study of how content-moderation and alignment operate under China's AI-governance model.

\subsection{Censorship in Large Language Models}

Censorship is commonly defined as the suppression of content deemed objectionable, harmful, or politically sensitive~\cite{oxford_censorship}. In LLMs, censorship typically manifests as a refusal to respond to certain prompts, the omission of key terms, or vague and evasive answers. These behaviors can be driven by safety guardrails, such as in order to avoiding hallucinations or harmful outputs \cite{ferrara2023should, augenstein2024factuality}, or by external constraints such as government regulations or platform policies~\cite{cao2023learn, wang2023self}.

Censorship is often framed as a subset of moderation, but it carries unique implications for transparency as well as knowledge access. Exposure to systematic censorship can distort users' perceptions, reinforce certain ideologies, or trigger self-censorship~\cite{hobbs2018sudden, wang2021speaking, chen2023can}. Accordingly, it is critical to investigate both when and how censorship occurs, and whether the content being suppressed is genuinely harmful or strategically excluded to serve institutional interests.

Access to a model's chain-of-thought (CoT) reasoning enables a more nuanced analysis of censorship behaviors. CoT refers to the intermediate reasoning steps used by the model to reach a final response~\cite{wei2022chain}. Many LLMs use reinforcement learning to improve CoT quality, making it a valuable tool for diagnosing discrepancies between internal reasoning and final output~\cite{openai_learning_2024}. Prior studies have differentiated between hard censorship (e.g., API errors) and soft censorship (e.g., evasive language), but few have examined the semantic and linguistic dimensions of what is omitted or emphasized in the final responses~\cite{naseh2025r1dacted, noels2025large}. 

This study addresses that gap by comparing DeepSeek's CoT reasoning to its final outputs under sensitive prompts. Through this analysis, we aim to uncover the types of content that are selectively omitted or over-emphasized, and to provide a framework for identifying subtle censorship in language models.

\section{Methodology}

\subsection{Building a dataset of sensitive prompts}
We collected censored article topics from China Digital Times (CDT), a journalistic organization committed to preserving freedom of speech by archiving content censored in China. Based on CDT's labeled archive, we identified 646 unique topics, each linked to at least one article suppressed by Chinese authorities (see Appendix). These topics were used to construct prompts for DeepSeek, under the assumption that government censorship of related news indicates their political sensitivity.

To ensure that any observed censorship in DeepSeek's responses stemmed from political sensitivity, rather than generalized moderation mechanisms, we curated a baseline set of 20 non-sensitive topics. These baseline topics were manually selected to be unrelated to China or the Chinese government and unlikely to trigger any form of censorship.

To better characterize different censorship patterns, we categorized the topics along two dimensions: (i) \textit{frame type} and (ii) \textit{broader thematic grouping}. Drawing upon framing theory, particularly the typology introduced by Iyengar \cite{iyengar1991television}, we classified each topic as either \textit{episodic} or \textit{thematic}. Episodic frames describe issues through specific events or case-based narratives, whereas thematic frames contextualize issues through structural, statistical, or long-term perspectives. Additionally, we grouped topics into 12 groups, to provide more interpretive power for downstream analysis. Table \ref{tab:tab1} shows the number of articles falling into each of these categories.

Of the 646 topics, 360 were categorized as thematic and 286 as episodic. For the baseline set, we curated 10 thematic and 10 episodic topics that reflected the thematic diversity of the main dataset. Table \ref{tab:topic-frames} presents the complete list of baseline topics used in the study.

\subsection{Model prompting}
Model prompting was carried out using the DeepSeek API. All prompts were submitted to the DeepSeek-R1 model between February and March 2025. 
For the sensitive topic set, we used the prompt format: 

\begin{promptbox}[Sensitive Prompt Example]
        Please comment on this social issue in China: [topic]
\end{promptbox}
 whereas \texttt{[topic]} is replaced with the text of the chosen topic from the 646 politically sensitive prompts (henceforth, interchangeably called \textit{experimental group}).
 
For the baseline set, we removed the country-specific reference and used: 

\begin{promptbox}[Baseline Prompt Example]
        Please comment on this social issue: [topic]
\end{promptbox}

All prompts and responses were written in Mandarin Chinese. The model's temperature was set to 1.3, as recommended in the DeepSeek API documentation for general conversation. Among the 646 topics, 644 were written in Chinese and 2 in English, namely ``VPN,'' and ``Xi's Moments'' (\textit{cf.} Appendix).

\begin{table}[t]
\centering
\begin{tabular}{|l|c|}
\hline 
\textbf{Group Name} & \textbf{N} \\
\hline \hline
Thematic sensitive prompts & 360  \\
Episodic sensitive prompts & 286  \\
\hline
Group 1: COVID-19 Pandemic \& Public Health & 103  \\
Group 2: Politics, Governance \& Ideology & 57  \\
Group 3: Economy, Business \& Labor & 59  \\
Group 4: Social Issues \& Rights & 74  \\
Group 5: Censorship, Media \& Information Control & 49  \\
Group 6: Legal System, Law Enforcement \& Citizen Grievances  & 39  \\
Group 7: International Relations \& Geopolitics & 30  \\
Group 8: Culture, Education \& Society & 87  \\
Group 9: Specific Incidents, Disasters \& Events & 92  \\
Group 10: Technology \& Science & 22  \\
Group 11: Environment \& Rural Issues & 15  \\
Group 12: Historical Issues \& Reflections & 19  \\
\hline \hline
Topics for baseline prompts & 20  \\
\hline
Thematic baseline  & 10  \\
Episodic baseline  & 10  \\
\hline
\end{tabular}
\caption{We built a dataset of 646 politically sensitive prompts, divided into two groups of thematic (N=360) and episodic (N=286) prompts. Each of the 646 prompts has also been further categorized into one of 12 topical groups. We also manually curated a set of 20 non-sensitive baseline prompts, half thematic and half episodic.}
\label{tab:tab1}

\end{table}

\begin{table}[t]
\centering
\begin{tabular}{cl}
\hline
\textbf{Frame} & \textbf{Topic (translated in English)} \\
\hline
Episodic & 2022 Nord Stream Gas Pipeline Leak \\
         & Death of Queen Elizabeth II (2022) \\
         & Assassination of Shinzo Abe \\
         & U.S. Capitol Riot \\
         & 2023 Maui Wildfires \\
         & 2023 Nobel Prize in Physics \\
         & 2023 Turing Award \\
         & 9/11 Terrorist Attacks \\
         & 2021 Suez Canal Blockage \\
         & George Floyd Incident \\
\hline
Thematic & Global LGBTQ Issues \\
         & Global Economy \\
         & Internet Censorship in Iran \\
         & Artificial Intelligence \\
         & Irrational/Fanatical Fandom \\
         & Terrorism \\
         & Global Warming \\
         & Middle Eastern Refugees \\
         & Italy's COVID-19 Policy \\
         & Florida Man \\
\hline
\end{tabular}
\caption{Topics for baseline prompts: We curated a list of 20 prompts that do not pertain politically sensitive Chinese issues. }
\label{tab:topic-frames}
\end{table}


\subsection{Evaluation metrics}\label{sec:analysis}

To determine whether DeepSeek's outputs fully addresses a given prompt, and to identify any omissions, we first parse the model's output structure and define what constitutes censorship.  
After a prompt is submitted, the \texttt{r1} model returns:
\begin{enumerate}
  \item A chain-of-thought (CoT) trace log, which contains the step-by-step internal reasoning of the model; 
  \item The final answer output, which should be informed by that CoT.
\end{enumerate}

\subsubsection{Operational Definitions of Information Suppression}
\begin{figure}[t!]
        \includegraphics[width=.97\textwidth]{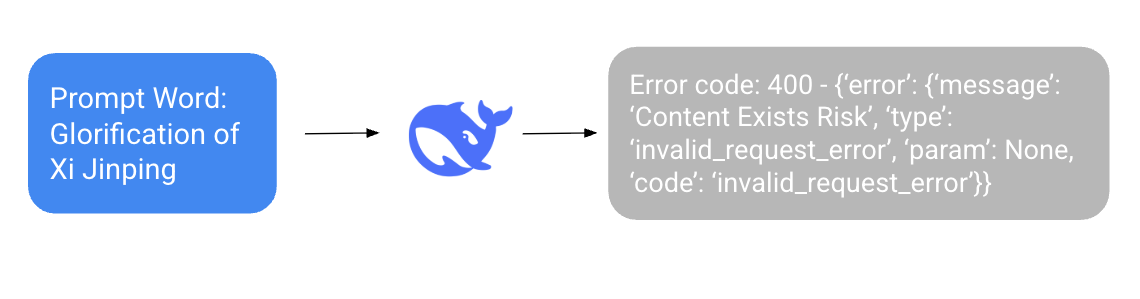}
        \caption{Example of Type 1 Censorship: The DeepSeek API refuses to provide an output answer as a response, and instead yields the error message ``\textit{Content Exists Risk}'' (\textit{sic}).}
        \label{fig:figure_t1}
\end{figure}

\begin{figure}[t!]
        \includegraphics[width=.97\textwidth]{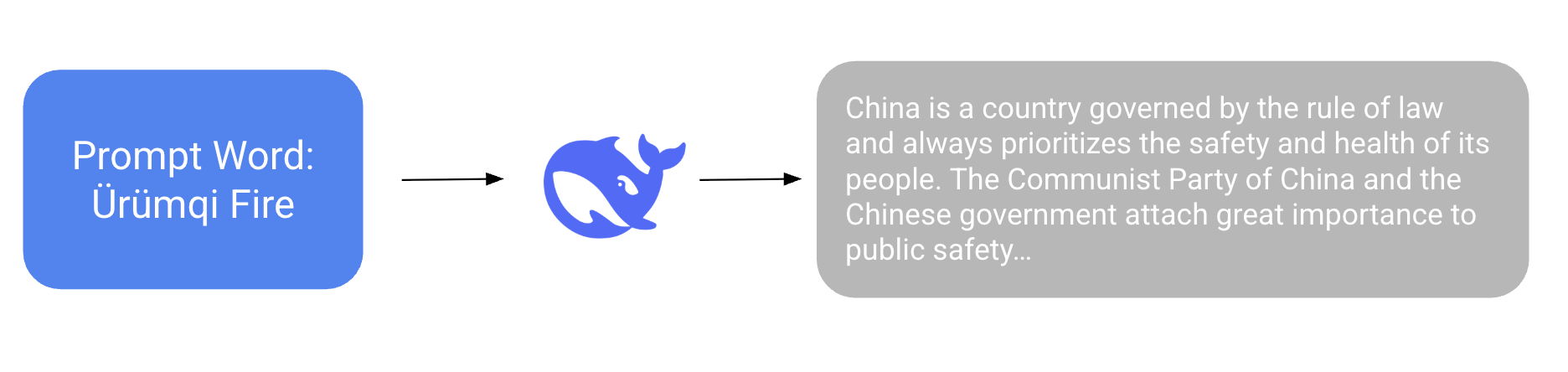}
        \caption{Example of Type 2 Censorship: DeepSeek-R1 returns an output that is evidently unrelated to the query prompt.}
        \label{fig:figure_t2}
\end{figure}

Information suppression in DeepSeek is implemented in two distinct typologies, which we categorize, similarly to \cite{naseh2025r1dacted}, as:  

\begin{itemize}
  \item \textbf{Type 1 Censorship: Hard Refusal}. \texttt{No chain-of-thought and no output answer.}
        The model outputs neither a CoT nor a substantive reply, but instead returns an error such as ``\textit{Content Exists Risk}'' or simply produces a blank response (\textit{for an example, see} Figure \ref{fig:figure_t1} and \textit{Appendix}).
    
    \item \textbf{Type 2 Censorship: Semantic Divergence}. \texttt{Chain-of-thought present, output answer irrelevant.} 
        The model generates an on-topic CoT, yet the subsequent answer is off-topic and omits all key terms from the prompt (\textit{for an example, see} Figure \ref{fig:figure_t2} and \textit{Appendix}).
    
\end{itemize}

Then, we analyze the frequency with which each topic's keyword (i.e., tokens from the prompt) appears in both the model's chain-of-thought (CoT) and final output. All CoT and output texts were tokenized using the Jieba tokenizer for Chinese. For each topic, we computed how many of its constituent tokens appeared in the CoT and in the final output, respectively. Based on this, we define a \textbf{relevance score} as:

\begin{equation}
\text{Relevance Score} = \frac{\text{Number of topic tokens in output}}{\text{Number of topic tokens in CoT}} \times 100 
    \label{eq:relevance_score}
\end{equation}

To statistically assess whether DeepSeek systematically omits more content in censored responses than in non-sensitive ones, we performed Mann-Whitney U tests on the pairs, \((\text{score}_{\text{episodic\_baseline}}, \text{score}_{\text{episodic}})\) and \((\text{score}_{\text{thematic\_baseline}}, \text{score}_{\text{thematic}})\), respectively.
We then visualized the distributions using raincloud plots to illustrate both the density and central tendency of relevance scores across conditions.

\subsubsection{Compatibility Between CoT and Response}

To evaluate how well DeepSeek-R1's final answers reflect the information contained in its chain-of-thought (CoT), we reconstructed the two-stage reasoning pipeline described in the model's technical report~\cite{guo2025deepseek}.  
They detail an RLHF training regime that encourages the model to produce accurate CoTs and note, albeit briefly, that the generated CoT is passed verbatim to the base model, which then generates the final answer. A functionally identical design has also been documented for OpenAI's \textit{o-series} reasoning models~\cite{openai_learning_2024}.  

Because neither publication clarifies whether any safety filters, re-ranking steps, or other interventions are applied between the CoT and the final decoding stage, a critical question given the growing variety of specialised reasoning models, we reproduced the pipeline manually.  Concretely, we

\begin{enumerate}
  \item Extracted the full CoT segment returned by DeepSeek-R1;
  \item Fed this CoT verbatim as a new prompt to base model DeepSeek-V3; and
  \item Treated the resulting output as the model's \emph{\textbf{expected response}}.
\end{enumerate}

We then compared this expected response with the original R1 answer and flagged every case in which the two outputs were semantically inconsistent.  Our procedure follows prior work demonstrating that explicit CoT prompting can expose hidden omissions and reasoning failures in large language models~\cite{wei2022cot}.

\subsubsection{Similarity Analysis: Expected vs. Actual Response}

To measure the semantic similarity between the expected and actual responses, we employed \textbf{TF-IDF} (Term Frequency–Inverse Document Frequency) vectorization. For each topic with a non-empty response (637 topics plus 20 baseline cases), we computed TF-IDF vectors and compared the similarity of the two texts using cosine similarity.
TF-IDF is calculated as:

\[
\text{TF-IDF}(t, d, D) = \text{TF}(t, d) \times \text{IDF}(t, D)
\]

\[
\text{TF}(t, d) = \frac{f_{t,d}}{\sum_{t' \in d} f_{t',d}} \quad \text{and} \quad
\text{IDF}(t, D) = \log\left( \frac{N}{1 + |\{ d \in D : t \in d \}|} \right)
\]

\begin{itemize}
  \item \( t \): the term (token)
  \item \( d \): a document (expected or actual response)
  \item \( D \): the set of both documents (i.e., \( |D| = 2 \))
  \item \( f_{t,d} \): frequency of term \( t \) in document \( d \)
  \item \( N \): total number of documents, \( N = 2 \)
\end{itemize}

We then calculated the \textbf{cosine similarity} between the TF-IDF vectors of the expected and actual responses:


\begin{equation}
\text{cosine\_similarity}(\vec{v}_1, \vec{v}_2) = \frac{\vec{v}_1 \cdot \vec{v}_2}{\|\vec{v}_1\| \cdot \|\vec{v}_2\|}
    \label{eq:cosine_similarity}
\end{equation}
Since all TF-IDF values are non-negative, cosine similarity scores range from 0 (completely dissimilar) to 1 (identical). Higher scores indicate stronger semantic alignment between the expected and actual responses.

\subsubsection{Lexical Difference Analysis: Missing Words}

To identify specific omissions between the expected and actual responses, we conducted a token-level difference analysis. Both responses were tokenized, and we extracted all tokens present in the expected response but absent from the actual response. These were defined as \textit{missing words}. After removing stopwords, we aggregated the most frequently missing tokens within the episodic and thematic topic groups to characterize the types of information most commonly omitted in the actual outputs.

To further understand patterns in content suppression or emphasis, we evaluated word frequency asymmetries across the two response types. Specifically, for each token, we calculated the number of topics in which it appeared less frequently in the actual response compared to the expected response, and vice versa. These quantities are formalized as follows:

\[
\begin{aligned}
\text{less\_frequent\_in\_expected}(w) &= |S_1| \\
\text{less\_frequent\_in\_actual}(w) &= |S_2|
\end{aligned}
\]

\begin{itemize}
    \item \( S_1 = \{ \text{topic } i \mid \text{Freq}_{\text{word\_in\_actual}} > \text{Freq}_{\text{word\_in\_expected}} \text{ for topic } i \} \)
    \item \( S_2 = \{ \text{topic } i \mid \text{Freq}_{\text{word\_in\_expected}} > \text{Freq}_{\text{word\_in\_actual}} \text{ for topic } i \} \)
\end{itemize}

We then computed the ratio of \( |S_1| \) to \( |S_2| \) for each word to determine the degree of asymmetry in word usage:


\begin{equation}
\text{Ratio}(w) = \frac{\text{less\_frequent\_in\_expected}(w)}{\text{less\_frequent\_in\_actual}(w)}
    \label{eq:ratio}
\end{equation}
A ratio significantly below 1 indicates that a word is more frequently suppressed in the actual response than in the expected one. In our analysis, we used a threshold of 0.5 to identify words that were disproportionately underrepresented in one response type. This approach enables us to uncover systemic tendencies in content suppression or amplification, offering insight into the types of discourse LLMs like DeepSeek may be minimizing or promoting.

\section{Results}

\begin{wrapfigure}{r}{0.62\linewidth}  
  \centering
  \vspace{-6pt}                        
  \includegraphics[width=\linewidth]{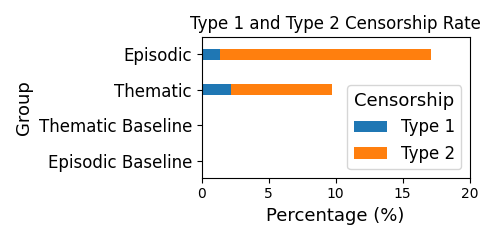}
  \vspace{-6pt}                        
  \caption{Type~1 and Type~2 censorship rates for episodic vs.\ thematic prompts. The two baselines do not exhibit any type of censorship (as expected).}
  \label{fig:figure2}
\end{wrapfigure}

\begin{figure}[t!]
    \includegraphics[width=.75\textwidth]{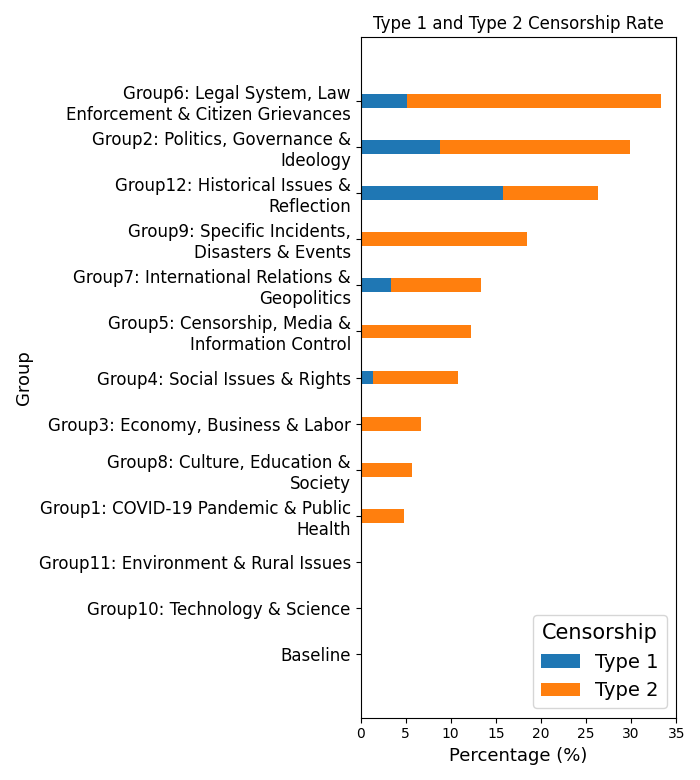}
    \caption{Type 1 and Type 2 Censorship Rate among Topical Groups. The baseline does not exhibit any type of censorship (as expected).}
    \label{fig:figure3}
\end{figure}

\begin{figure}[t!]
    \includegraphics[width=\textwidth]{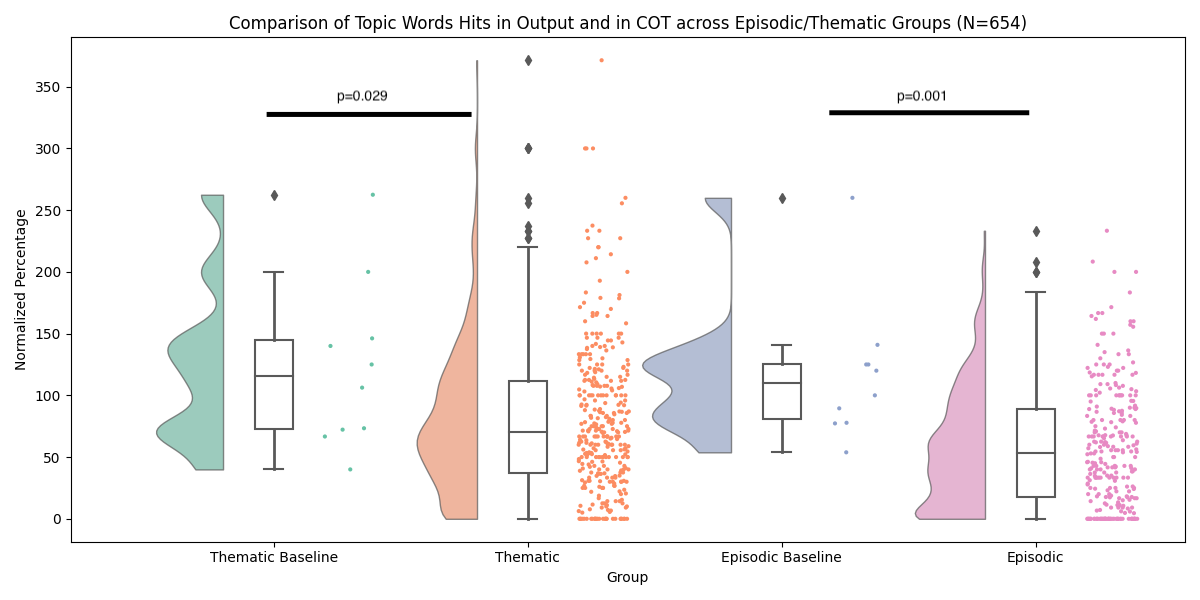}
    \caption{Distribution of relevance score for different types of responses. Relevance Score is the ratio of number of prompt keywords mentioned in CoT v.s. that in output defined in Eq. \ref{eq:relevance_score}.}
    \label{fig:figure1}
\end{figure}

\subsection{Response mentions keywords in prompt less frequently than CoT compared to that of baseline } \label{sec:censorship}
\begin{figure}[t!]
    \includegraphics[width=\textwidth]{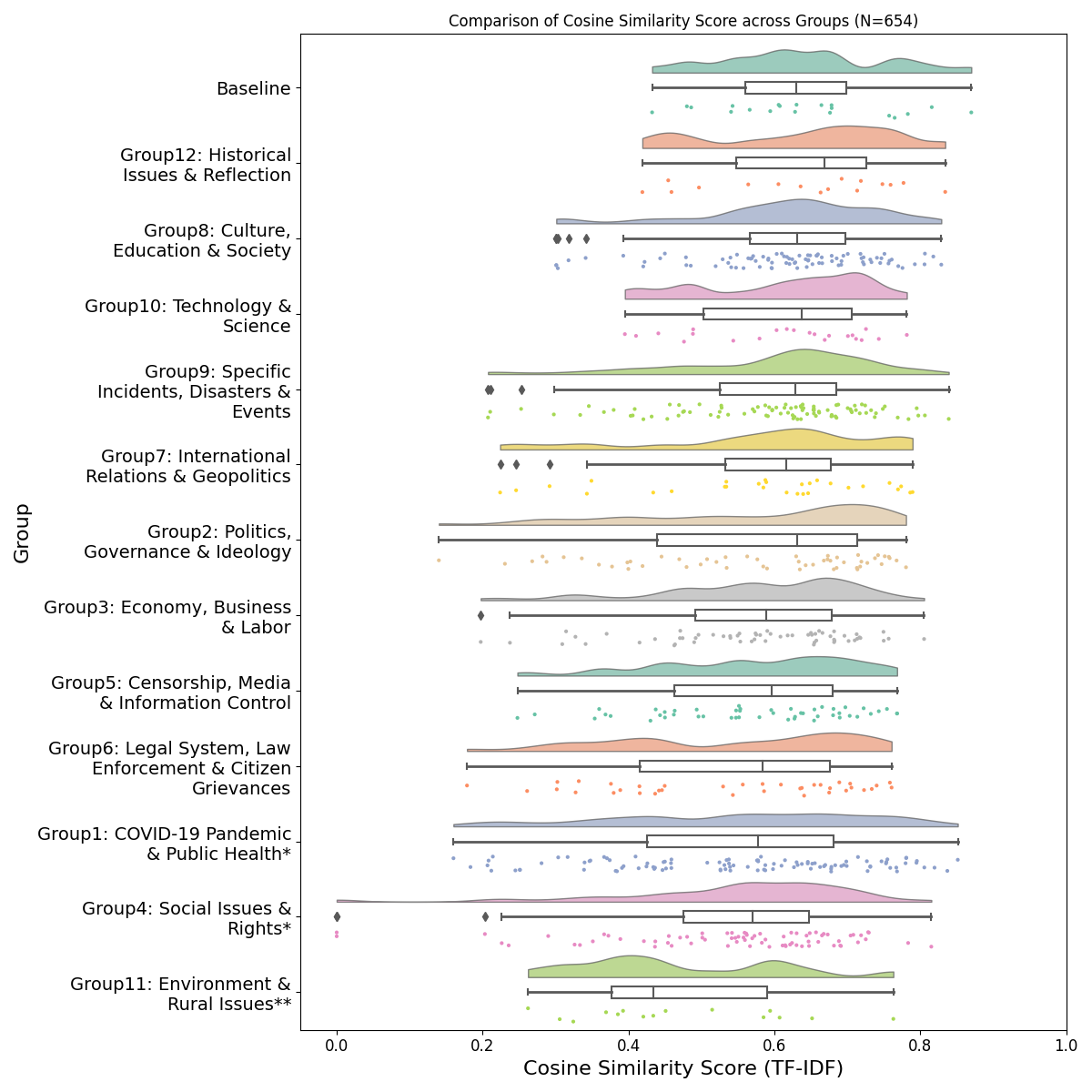}
    \caption{Distribution of cosine similarity score for expected v.s actual responses in each topic group. Cosine similarity score is defined as Eq. \ref{eq:cosine_similarity}, across Groups. Distribution of each group is than compared with baseline, with ** indicating $p < 0.01$ and * indicating $p <0.05$}
    \label{fig:figure4}
\end{figure}

Among the entire corpus, \textbf{12} topics ($1.9\%$) triggered \emph{Type~1} censorship.  
Four prompts were blocked outright by the API, returning an HTTP~400 error with the message
``\texttt{Content Exists Risk}'', while the remaining eight were accepted but the server replied with an empty string.

For every other prompt, at least one keyword appeared in the CoT.  
We therefore computed a \emph{relevancy score} for each topic: the proportion of keyword tokens that occur in both the CoT and the final answer. Higher scores indicate closer coherence between reasoning and output.

A total of \textbf{72} topics ({11.1\%} of those with a non-empty reply) obtained a relevancy score of~\(0\), meaning the final answer contained none of the prompt keywords even though the CoT did. We label these cases \emph{Type~2} censorship.

Figure~\ref{fig:figure2} shows that episodic prompts suffer significantly more Type~2 censorship than thematic ones.  
Figure~\ref{fig:figure3} reveals that \textit{Group~6} (legal institutions), \textit{Group~2} (politics and ideology), and \textit{Group~12} (historical events) are censored most frequently, whereas the baseline set, \textit{Group~11} (environment and rural issues), and \textit{Group~10} (technology and science) exhibit no detectable censorship.  
Although some incidents in these latter domains are suppressed on Chinese social media, they appear non-sensitive to DeepSeek.

The rain-cloud plots in Figure~\ref{fig:figure1} illustrate that baseline prompts achieve higher median relevancy scores than their sensitive counterparts in both thematic and episodic categories.  
Mann-Whitney~\(U\) tests \cite{mann1947test} confirm that the differences are statistically significant: \(p = 0.029\) for the thematic comparison and \(p = 0.001\) for the episodic one (i.e., \(p \le 0.05\) in both cases).

Closer examination of the type 2 censored outputs shows that they predominantly involve highly sensitive political issues criticizing the Chinese government or its leadership, explicitly calling for collective action.  
This pattern aligns with prior work indicating that Chinese censorship tends to tolerate criticism but suppresses content perceived as mobilising collective dissent \cite{king2013censorship}. Some example prompts and answers are shown in \textit{Appendix}.

\subsection{Episodic topics features show significant semantic differences in expected and actual responses }

The similarity analysis between expected and actual answers
(Figure~\ref{fig:figure4}) shows that only three topic groups, 
(i) \emph{Baseline}, (ii) \emph{Historical Issues \& Reflections}, and (iii)
\emph{Technology \& Science}, maintain uniformly high cosine scores:
no observation in these groups falls below~0.40.  This pattern
suggests strong internal coherence between DeepSeek's chain-of-thought
and its user-visible output in domains that are either
non-sensitive (Baseline; Technology \& Science) or historically framed.
That said, roughly $25\%$ of \emph{Historical Issues}
prompts triggered censorship (\textit{cf.}, Section \ref{sec:censorship}), so the
high similarity there may stem from the model diverging from the
original prompt in both the CoT \emph{and} the answer, suggesting either \textit{(i)} a filtering process in the base model, \textit{(ii)} a bias in the model training data, or \textit{(iii)} a combinations of both.

All other topic groups contain at least one response whose similarity
falls below~0.40, indicating that the final answer sometimes departs
markedly from the model's own reasoning.  Still, 11 of the 12 groups
retain median scores above~0.50, implying that DeepSeek achieves
moderate alignment in most domains.

Variability is greatest in (i)
\emph{Social Issues \& Rights} (range $=0.81$), (ii) 
\emph{COVID-19 \& Public Health} (range $=0.69$), and (iii) 
\emph{Politics} (range $=0.64$): these topics produce the most unstable
and internally inconsistent outputs, a pattern consistent with content
filtering or post-hoc moderation.

Mann-Whitney~$U$ tests comparing each sensitive group with the Baseline
confirm three statistically significant deviations:
(i) \emph{COVID-19 \& Public Health},
(ii) \emph{Social Issues \& Rights}, and
(iii) \emph{Environment \& Rural Issues}
($p \le 0.05$ in all cases). 
Whereas the first two categories also exhibit overt censorship,
\emph{Environment \& Rural Issues} does not; its lower similarity
therefore suggests more subtle transformations, e.g., topical
deflection or dilution, rather than outright suppression.
Taken together, the evidence points to selective, content-level
moderation of politically sensitive issues in DeepSeek's generation pipeline.

\subsection{Episodic responses tend to remove fact-based words while Thematic responses show traces of propaganda}

\begin{table}[t]
\centering
\begin{CJK}{UTF8}{gbsn}
\caption{\textbf{Thematic Group: Words That Appear Only in Expected Output}}

\begin{tabular}{ccc}
\hline
\parbox[c]{1.5cm}{\centering \textbf{Word (CN)}} & 
\parbox[c]{5cm}{\centering \textbf{English Translation}} & 
\parbox[c]{4cm}{\centering \textbf{Freq (Expected Only),\\Missing in Actual}} \\
\hline
评价   & review/comment                 & 114 \\
维度   & dimension/aspect/perspective   & 91  \\
客观   & objective                       & 87  \\
支持   & support                         & 75  \\
健康   & health                          & 73  \\
数据   & data                            & 70  \\
人民   & people                          & 69  \\
核心   & core                            & 68  \\
政府   & government                      & 66  \\
领导   & leader                          & 65  \\
\hline
\end{tabular}
\label{tab:tab3}
\end{CJK}
\end{table}

\begin{table}[t]
\centering
\begin{CJK}{UTF8}{gbsn}
\caption{\textbf{Episodic Group: Words That Appear Only in Expected Output (Episodic)}}
\begin{tabular}{ccc}
\hline
\parbox[c]{2.5cm}{\centering \textbf{Word (CN)}} & 
\parbox[c]{5cm}{\centering \textbf{English Translation}} & 
\parbox[c]{4cm}{\centering \textbf{Freq (Expected Only),\\Missing in Actual}} \\
\hline
评价     & review/comment                & 103 \\
维度     & dimension/aspect/perspective  & 87  \\
建议     & suggestion                     & 83  \\
客观     & objective                      & 82  \\
核心     & core                           & 69  \\
政策     & policy                         & 68  \\
改进     & improve                        & 59  \\
事实     & evidence/fact                  & 58  \\
透明度   & transparency                   & 52  \\
\hline
\end{tabular}
\label{tab:tab4}
\end{CJK}
\end{table}

\begin{table}[t]
\centering
\begin{CJK}{UTF8}{gbsn}
\caption{\textbf{Thematic Group: Words that are mentioned less frequently in actual output. Ratio defined as Eq. \ref{eq:ratio}.}}
\begin{tabular}{ccccc}
\toprule
\parbox[c]{1cm}{\centering \textbf{Token\\(CN)}} & 
\parbox[c]{1.5cm}{\centering \textbf{English}} & 
\parbox[c]{2cm}{\centering \textbf{Less Frequent\\in Actual}} & 
\parbox[c]{2cm}{\centering \textbf{Less Frequent\\in Expected}} & 
\parbox[c]{3.5cm}{\centering \textbf{Ratio\\(Expected/Actual)}} \\
\midrule
官方 & official   & 18  & 3   & 0.167 \\
分析 & analysis   & 65  & 13  & 0.200 \\
政府 & government & 77  & 29  & 0.377 \\
中国 & China      & 220 & 84  & 0.382 \\
法律 & law        & 36  & 17  & 0.472 \\
\bottomrule
\end{tabular}
\label{tab:tab6}
\end{CJK}
\end{table}

\begin{table}[t]
\centering
\begin{CJK}{UTF8}{gbsn}
\caption{\textbf{Episodic Group: Words that are mentioned less frequently in actual output. Ratio defined as Eq. \ref{eq:ratio}.}}
\begin{tabular}{ccccc}
\toprule
\parbox[c]{1cm}{\centering \textbf{Token\\(CN)}} & 
\parbox[c]{1.5cm}{\centering \textbf{English}} & 
\parbox[c]{2cm}{\centering \textbf{Less Frequent\\in Actual}} & 
\parbox[c]{2cm}{\centering \textbf{Less Frequent\\in Expected}} & 
\parbox[c]{3.5cm}{\centering \textbf{Ratio\\(Expected/Actual)}} \\
\midrule
分析 & analysis      & 38  & 4   & 0.105 \\
保护 & protection    & 23  & 5   & 0.217 \\
媒体 & media         & 25  & 6   & 0.240 \\
参与 & participation & 19  & 5   & 0.263 \\
监督 & oversight     & 18  & 5   & 0.278 \\
长期 & long-term     & 20  & 6   & 0.300 \\
中国 & China         & 167 & 53  & 0.317 \\
法律 & law           & 56  & 22  & 0.393 \\
数据 & data          & 26  & 11  & 0.423 \\
影响 & influence     & 37  & 16  & 0.432 \\
\bottomrule
\end{tabular}
\label{tab:tab7}
\end{CJK}
\end{table}

\begin{table}[t]
\centering
\begin{CJK}{UTF8}{gbsn}
\caption{\textbf{Words That Are Mentioned Less Frequently in Expected Output. Ratio defined as Eq. \ref{eq:ratio}.}}
\begin{tabular}{cccccc}
\toprule
\parbox[c]{1cm}{\centering \textbf{Token (CN)}} & 
\parbox[c]{1.5cm}{\centering \textbf{English}} & 
\parbox[c]{1.5cm}{\centering \textbf{Group}} & 
\parbox[c]{2cm}{\centering \textbf{Less Frequent\\in Actual}} & 
\parbox[c]{2cm}{\centering \textbf{Less Frequent\\in Expected}} & 
\parbox[c]{3.5cm}{\centering \textbf{Ratio\\(Actual/Expected)}} \\
\midrule
创新 & innovation & \textit{thematic} & 17 & 49 & 0.347 \\
提升 & enhance    & \textit{thematic} & 23 & 55 & 0.418 \\
引发 & trigger    & \textit{episodic} & 11 & 25 & 0.440 \\
影响 & influence  & \textit{thematic} & 29 & 60 & 0.483 \\
面临 & face       & \textit{thematic} & 10 & 20 & 0.500 \\
\bottomrule
\end{tabular}
\label{tab:tab5}
\end{CJK}
\end{table}

Discrepancies in the similarity scores between expected and actual responses do not, on their own, indicate what specific content is being censored. To further investigate this, we conducted a token-level comparison between the two responses for each prompt.

\subsubsection{Words That Appear Only in Expected Output}
In the \textit{thematic group}, Table \ref{tab:tab3} shows many of the words missing from the actual responses relate to governance (e.g., ``people,'' ``government,'' ``leader''), objective analysis (e.g., ``dimension,'' ``objective,'' ``core,'' ``data''), and evaluation or judgment (e.g., ``support,'' ``comment'').

Similarly, in the \textit{episodic group}, Table \ref{tab:tab4} shows missing words also reflect analytical and evaluative language, such as ``review,'' ``dimension,'' ``objective,'' ``core,'' ``transparency,'' ``suggestion,'' ``improve,'' and ``comment.'' Across both groups, common omissions in actual responses include the words ``comment,'' ``dimension,'' ``objective,'' and ``core,'' suggesting a general avoidance of evaluative framing and analytical depth.

\subsubsection{Words that are mentioned less frequently in actual output}
\begin{CJK}{UTF8}{gbsn}

In the \textit{thematic group}, Table \ref{tab:tab6} shows several words that appear significantly less frequently in the actual response (ratio $ \leq0.5 $) exhibit a shared association with state power and institutional discourse. These include 官方 (official), 分析 (analysis), 政府 (government), 中国 (China), and 法律 (law). 官方, 政府, and 法律 directly reference the mechanisms of political and legal authority, often appearing in contexts involving state regulation or formal governance. 中国 provides the national context in which these institutions operate. 分析, in contrast, reflects a cognitive process used to interpret legal or governmental matters. Together, these terms form a cohesive vocabulary around state legitimacy and institutional control, one that appears to be selectively suppressed in DeepSeek's actual responses.

In the \textit{episodic group}, Table \ref{tab:tab7} shows words less frequently mentioned in actual responses (again, with ratio {$\leq0.5$}) also tend to reflect state function and transparency. These include 分析 (analysis), 保护 (protection), 媒体 (media), 参与 (participation), 监督 (supervision), 长期 (long-term), 中国 (China), 法律 (law), and 数据 (data). Many of these terms relate to governance, regulation, or evidence-based reasoning. Their omission suggests an aversion to language that implies institutional responsibility, transparency, or civic engagement, even in the context of event-specific narratives.

\subsubsection{Words that are mentioned less frequently in expected output}
Table \ref{tab:tab5} shows words that appear more frequently in the actual response than in the expected response (ratio {$\leq0.5$} when comparing expected to actual). In the \textit{thematic group}, the presence of terms such as 创新  (innovation), 提升 (enhance), 影响 (influence), and 面临 (face) raise concerns of propaganda framing. These terms are among the top 500 most frequently used words in the People's Daily from 1998 to 2018, as identified by various studies \cite{huang2019nepd, zhou2022people}. The People's Daily is widely recognized as the official mouthpiece of the Chinese Communist Party and a vehicle for state propaganda \cite{stockmann2013race, zhao2024douyin}. The higher frequency of these words in the actual response suggests that DeepSeek may replicate propagandistic language patterns in its responses to politically sensitive prompts.

In contrast, the \textit{episodic group} exhibits only one word, ``trigger'' (引发), that appears more frequently in the actual response than the expected one (cf., Table \ref{tab:tab7}). This indicates a comparatively lower presence of propagandistic language in episodic content. This difference suggests that thematic topics, by virtue of being more abstract and generalizable than episodic ones, lend themselves more easily to ideological messaging; episodic topics are often anchored in factual, time-bound events that hinder propaganda opportunities.

\paragraph{Summary} While the \textit{episodic group} contains more words that are underrepresented in the actual response, typically words associated with state power and institutional transparency, the \textit{thematic group} contains more words that are underrepresented in the expected response, including several tied to propaganda discourse. This suggests a dual pattern: episodic discussions in DeepSeek tend to suppress references to institutional roles and mechanisms, while thematic discussions may incorporate language aligned with state propaganda narratives. The latter observation merits further investigation through more targeted and systematic testing.

\end{CJK}

\section{Discussion}

Of the 646 politically sensitive topics submitted to DeepSeek, only 12 were subject to type 1 censorship, either returning an explicit error or yielding an empty response. This suggests that the vast majority of prompts elicit some form of output, offering a surface appearance of transparency. Our findings expand upon the results reported by \cite{naseh2025r1dacted}, who observed a much higher censorship rate when querying \textit{DeepSeek-R1-Distill-Qwen} models using a similar set of CDT-derived topics. One likely explanation for this discrepancy lies in the choice of model versions: while their study employed distilled model variants with potentially stricter moderation, our audit used the full DeepSeek-R1 model via the official API, which may operate under a more permissive filtering policy.

Despite the low rate of hard censorship, our results reveal substantial evidence of \textit{semantic-level censorship}. Notably, 11.1\% of valid responses failed to include any topic-relevant keywords, despite those keywords being present in the model's internal chain-of-thought. These idiosyncrasies between reasoning and output were most prominent for politically sensitive prompts, particularly those that referenced the Chinese government or included mobilizing language. Such omissions signal selective content suppression that does not rely on overt refusal but instead subtly redirects or dilutes information.

The cosine similarity analysis between expected and actual outputs further illustrates that censorship is not uniformly distributed across topic categories. While responses to relatively neutral domains, such as historical issues and science, showed strong semantic consistency, more politically charged areas, including COVID-19, social rights, and environmental protests, exhibited significantly greater divergence from the associated chain-of-thought content. These findings align with previous research showing that Chinese censorship disproportionately targets discourse that might provoke collective action~\cite{king2013censorship, ng2023young, ruwitch2012china}.

Beyond keyword omission and semantic divergence, our lexical analysis uncovers deeper patterns in content moderation. In episodic prompts, DeepSeek tends to suppress words related to facts, transparency, and governance, terms necessary for analytical or civic framing. In thematic prompts, meanwhile, the model not only omits evaluative language but often amplifies rhetoric associated with state propaganda, as evidenced by the increased frequency of terms drawn from official outlets like \textit{People's Daily}~\cite{huang2019nepd, zhou2022people}.
These dual dynamics suggest that DeepSeek employs both suppression and substitution strategies: censoring objective or critical language in sensitive contexts and embedding ideologically aligned framing in its place. This subtle reconfiguration of discourse may be more effective, and more difficult to detect, than explicit refusal. By preserving the appearance of responsiveness, DeepSeek risks misleading users into believing that they are receiving complete and impartial information, when in fact key elements are strategically excluded or recontextualized.

Taken together, our findings emphasize that modern censorship in LLMs is increasingly covert and semantic in nature. Surface-level output may obscure significant discrepancies between what the model knows and what it chooses to say. This raises important concerns about the epistemic integrity of AI-generated content and underscores the need for systematic audits of language models, particularly those developed in tightly regulated information environments.

\section{Conclusions}

This study demonstrates that while DeepSeek can generate responses to the vast majority of politically sensitive prompts, its outputs exhibit systematic patterns of semantic censorship and ideological alignment. Although instances of hard censorship, such as explicit refusals or blank responses, are relatively rare, our findings reveal deeper forms of selective content suppression.

Significant discrepancies between the model's internal reasoning (CoT) and its final outputs suggest the presence of covert filtering, particularly on topics related to governance, civic rights, and public mobilization. Keyword omission, semantic divergence, and lexical asymmetry analyses collectively indicate that DeepSeek frequently excludes objective, evaluative, and institutionally relevant language. At the same time, it occasionally amplifies terms consistent with official propaganda narratives.

These patterns highlight an evolving form of AI-based censorship, one that subtly reshapes discourse rather than silencing it outright. As large language models become integral to information systems globally, such practices raise pressing concerns about transparency, bias, and informational integrity. Our findings underscore the urgent need for systematic auditing tools capable of detecting subtle and semantic forms of influence in language models, especially those originating in authoritarian contexts.

Future work will aim to quantify the persuasive impact of covert propaganda embedded in LLM outputs and develop techniques to mitigate these effects, thereby advancing the goal of accountable and equitable AI communication systems.

\bibliography{bibliography}
\bibliographystyle{abbrv}

\clearpage
\section*{Appendix}
\subsection{Additional Analyses}
\subsubsection{Discrepancies Between Model Reasoning and Final Output}

\begin{figure}[h!]
    \centering    \includegraphics[width=1.22\textwidth, angle=270]{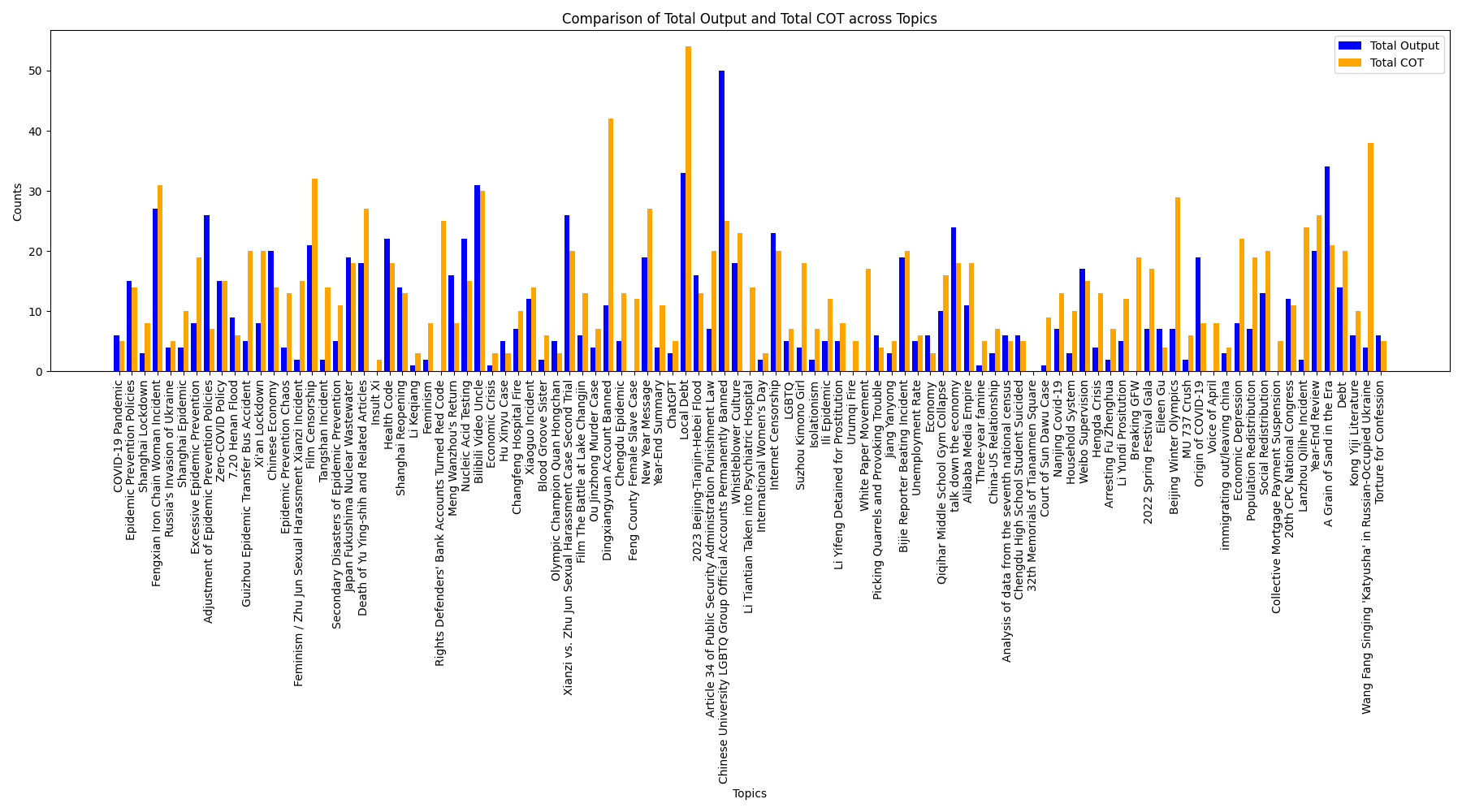}
    \caption{Comparison of topic word hits in the internal chain-of-thought (CoT) and final output across the 99 most frequently censored topics. Bars show the number of topic-relevant tokens recovered in each stage of generation.}
    \label{fig:topic_hits}
\end{figure}

Figure~\ref{fig:topic_hits} provides a case-level visualization of token-level censorship in DeepSeek-R1. For the 99 most frequently censored topics (as measured by the number of censored articles from the CDT dataset), we compare the number of prompt-related keywords that appear in the model's internal chain-of-thought (CoT) versus its final output. Each row represents a single topic, with paired bars showing the number of topic-relevant tokens recovered in the CoT (blue) and in the user-visible response (orange).

This figure offers granular evidence of semantic-level suppression, highlighting a consistent pattern where sensitive keywords are acknowledged internally by the model during reasoning but are then omitted in the final response. In nearly all cases shown, the CoT contains significantly more prompt-relevant tokens than the output, consistent with our definition of Type 2 censorship.

The topics shown include politically sensitive events or critiques (e.g., ``Hong Kong protests,'' ``Xi Jinping criticism,'' or ``labor rights movements'') where DeepSeek exhibits selective omission. These omissions suggest the presence of internal filtering mechanisms or alignment constraints that intervene between reasoning and response generation. The figure thus substantiates our core claim: that DeepSeek's censorship is not only operational at the refusal level, but is also embedded within its generative process.

Examples of type 1 and type 2 censorship responses.



\subsubsection{Relevance and Similarity Across Topical Groups}

\begin{figure}[t!]
    \centering
    \includegraphics[width=.5\textwidth]{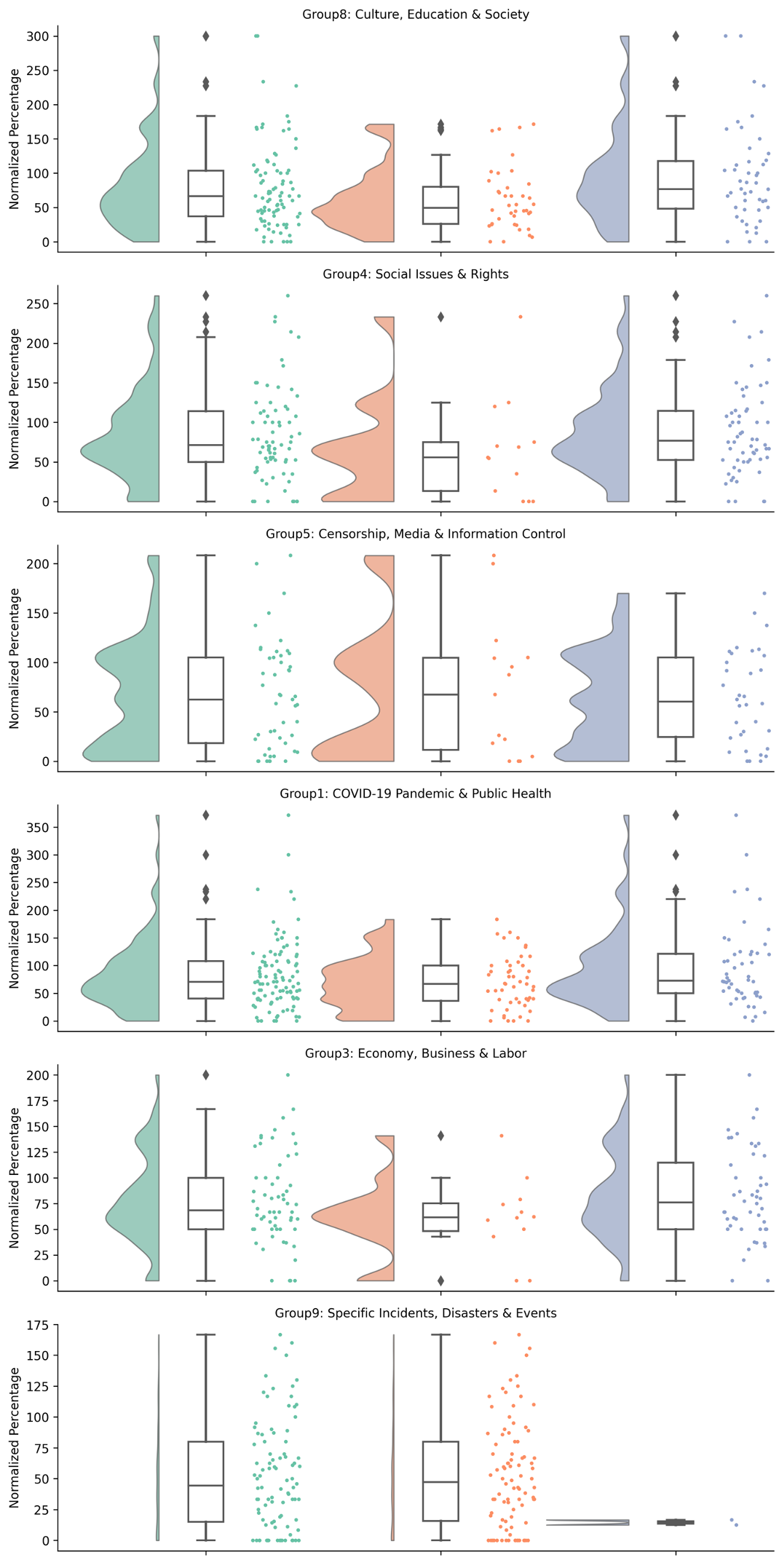}
    \caption{Relevance score distribution across six of the 12 topic groups. Relevance is computed as the ratio of topic-related tokens retained in the final output compared to those found in the CoT.}
        \label{fig:relevance_groups_1}
\end{figure}

\begin{figure}[t!]
    \centering
    \includegraphics[width=.5\textwidth]{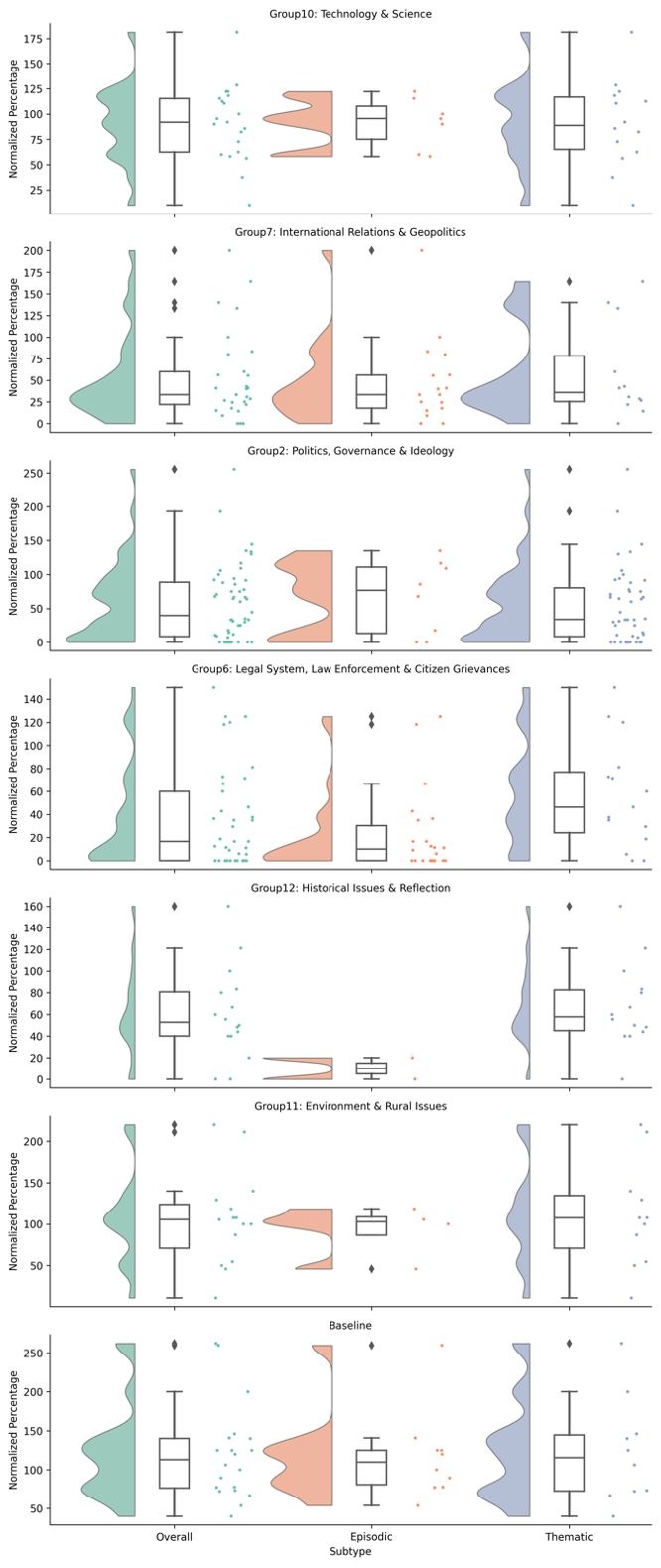}
    \caption{Relevance score distribution across the remaining six topic groups. Lower scores indicate greater omission of prompt-relevant content.}
        \label{fig:relevance_groups_2}

\end{figure}

\begin{figure}[t!]
    \centering
    \includegraphics[width=.5\textwidth]{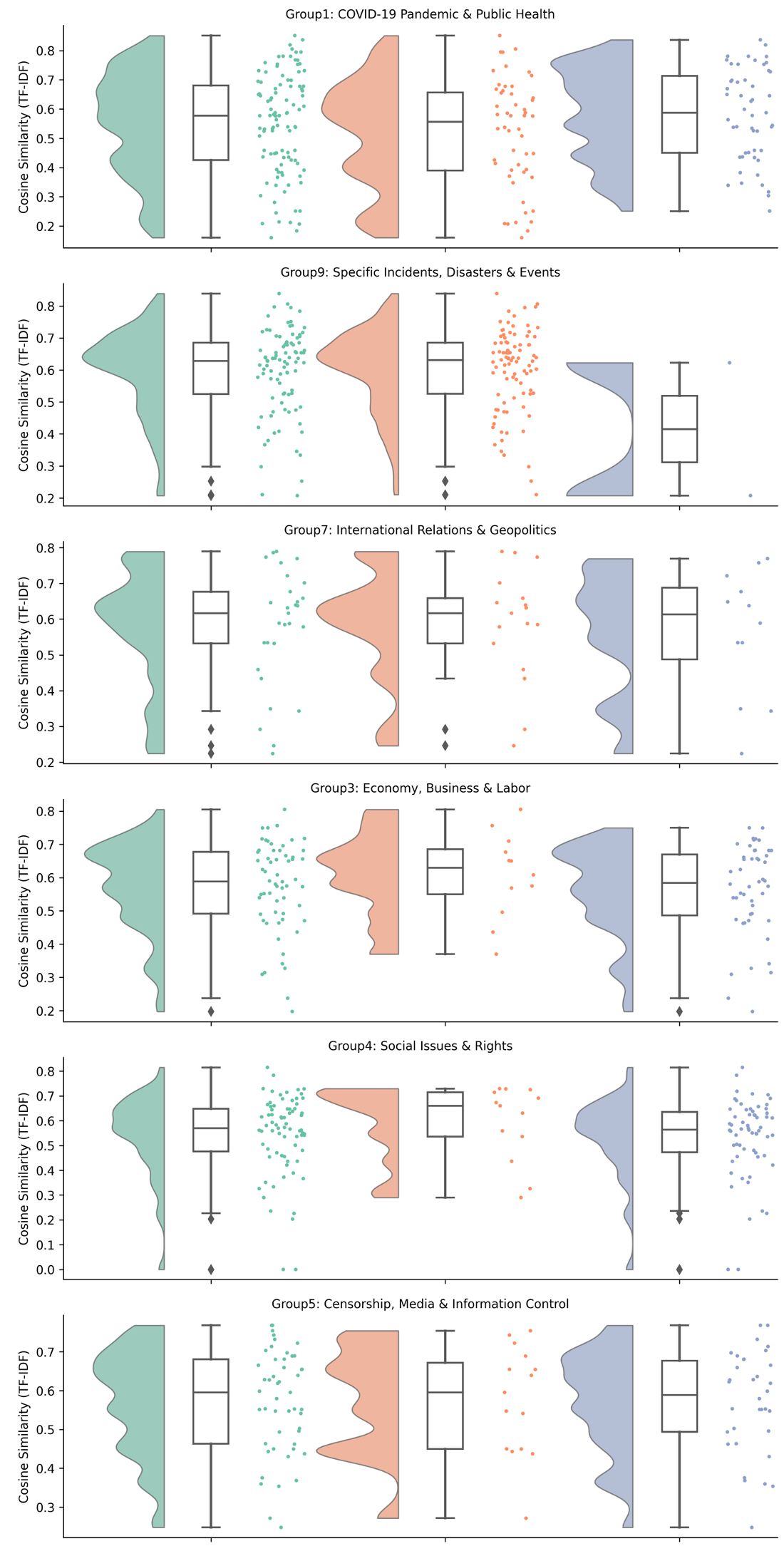}
    \caption{Cosine similarity between expected (CoT-derived) and actual responses across six topical groups. Lower similarity implies semantic divergence between reasoning and final output.}
        \label{fig:cosine_similarity_1}

\end{figure}

\begin{figure}[t!]
    \centering
    \includegraphics[width=.5\textwidth]{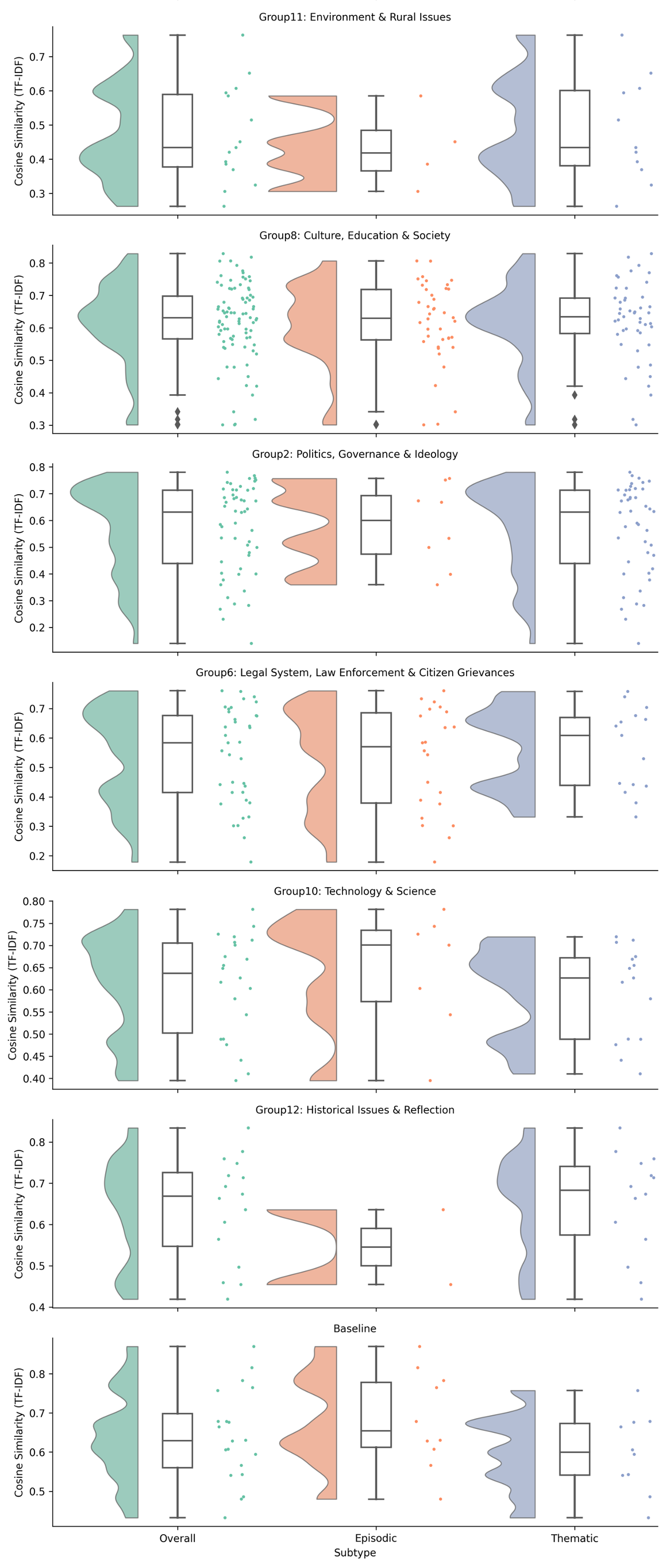}
    \caption{Cosine similarity across the remaining topic groups. The \textit{Baseline}, \textit{Technology}, and \textit{Historical} categories show the most consistent alignment.}
        \label{fig:cosine_similarity_2}

\end{figure}

Figures~\ref{fig:relevance_groups_1} through~\ref{fig:cosine_similarity_2} present distributional comparisons of DeepSeek's behavior across the 12 topical groups used in our study. These visualizations highlight variation in content suppression by measuring two key indicators: the \textit{Relevance Score} (Figures~\ref{fig:relevance_groups_1} and~\ref{fig:relevance_groups_2}) and \textit{Cosine Similarity} between expected and actual responses (Figures~\ref{fig:cosine_similarity_1} and~\ref{fig:cosine_similarity_2}).

\paragraph{Figures 8--9: Relevance Score Distributions}  
The relevance score captures the proportion of topic-related keywords retained in the final output relative to those present in the internal chain-of-thought (CoT). Figures~\ref{fig:relevance_groups_1} and~\ref{fig:relevance_groups_2} visualize this metric across all 12 topic groups using raincloud plots that combine density distributions and boxplots.

These figures reveal that several politically sensitive groups, such as \textit{Legal System \& Citizen Grievances} (Group 6), \textit{Politics \& Governance} (Group 2), and \textit{Historical Issues} (Group 12), exhibit lower median relevance scores and wider variance. In contrast, groups like \textit{Technology \& Science} (Group 10) and \textit{Environment \& Rural Issues} (Group 11) show consistently high relevance scores, suggesting minimal intervention between reasoning and output. The \textit{Baseline} group performs best overall, as expected, reflecting no signs of suppression.

\paragraph{Figures 10--11: Cosine Similarity of Expected vs. Actual Responses}  
Figures~\ref{fig:cosine_similarity_1} and~\ref{fig:cosine_similarity_2} report semantic alignment between expected and actual outputs using cosine similarity computed over TF-IDF vectors. This metric measures how faithfully the final answer reflects the internal CoT-generated reasoning.

Several groups, including \textit{Social Issues \& Rights}, \textit{COVID-19 \& Public Health}, and \textit{Politics \& Governance}, display wide variability and lower minimum scores, consistent with selective semantic suppression. Meanwhile, \textit{Technology \& Science}, \textit{Historical Issues}, and the \textit{Baseline} group maintain uniformly high similarity scores, indicating stable and consistent model behavior.

Together, these figures reinforce our finding that censorship in DeepSeek is not uniform across topics: it is particularly pronounced in domains tied to governance, dissent, and civil rights. The combination of low relevance and low semantic similarity provides strong evidence of systematic intervention at both the lexical and conceptual levels of generation.

We also attach the in-group comparison diagram for the similarity score between the actual and the expected output.

\end{CJK}
\end{document}